\documentclass[a4paper,10pt,twoside]{cpc-hepnp}
\usepackage{multicol}
\usepackage{graphicx}
\usepackage{booktabs}
\usepackage{amssymb,bm,mathrsfs,bbm,amscd}
\usepackage[tbtags]{amsmath}
\usepackage{lastpage}
\usepackage{subcaption}
\usepackage{hepunits}
\usepackage{lineno,hyperref}

\newif\ifpapercjk
\newif\ifpaperpublic
\newif\ifpaperthanks
\papercjkfalse
\paperpublicfalse
\paperthanksfalse
\ifpapercjk
\usepackage{CJK}
\fi
\modulolinenumbers[1]

\newcommand{\sep}{, }

\begin{document}
\ifpapercjk
\begin{CJK*}{UTF8}{gbsn}
\fi

% header and foot
\ifpaperpublic
    \fancyhead[c]{\small Chinese Physics C~~~Vol. XX, No. X (201X)
    XXXXXX} \fancyfoot[C]{\small 010201-\thepage}
    %\footnotetext[0]{Received 14 February 201X}
    \footnotetext[0]{Received DD MM 201X}
\else
    \fancyhead[c]{\small Submitted to `Chinese Physics C'} 
    \fancyfoot[C]{\small \thepage}
\fi

\title{Fast Muon Simulation in the JUNO Central Detector
    \thanks{This work is supported by the Strategic Priority Research Program of the Chinese Academy of Sciences (Grant No. XDA10010900),
            National Natural Science Foundation of China (Grant No. 11405279, 11575224).
            } 
}

\newcommand{\IHEP}{1}
\newcommand{\SYSU}{2}
\newcommand{\NKU}{3}
\author{
            Lin Tao \ifpapercjk(林韬)\fi$^{\IHEP;1}$\email{lintao@ihep.ac.cn}
      \quad Deng Zi-Yan \ifpapercjk(邓子艳)\fi$^{\IHEP}$
      \quad Li Wei-Dong \ifpapercjk(李卫东)\fi$^{\IHEP}$ 
      \quad Cao Guo-Fu \ifpapercjk(曹国富)\fi$^{\IHEP}$
      \\
      \quad You Zheng-Yun \ifpapercjk(尤郑昀)\fi$^{\SYSU}$
      \quad Li Xin-Ying \ifpapercjk(李新颖)\fi$^{\NKU}$
}
\maketitle
\address{
    $^{\IHEP}$ Institute of High Energy Physics, Chinese Academy of Sciences, Beijing 100049, China \\
    $^{\SYSU}$ School of Physics, Sun Yat-sen University, Guangzhou 510275, China \\
    $^{\NKU}$ School of Physics, Nankai University, Tianjin 300071, China\\
}

% == abstract ==
\begin{abstract}
    The Jiangmen Underground Neutrino Observatory (JUNO) is a multi-purpose neutrino experiment designed to measure the neutrino mass hierarchy using a central detector (CD), which contains 20 kton liquid scintillator (LS) surrounded by about 17,000 photomultiplier tubes (PMTs). Due to the large fiducial volume and huge number of PMTs, the simulation of a muon particle passing through the CD with the Geant4 toolkit becomes an extremely computation-intensive task. This paper presents a fast simulation implementation using a so-called voxel method: for scintillation photons generated in a certain LS voxel,  the PMT's response is produced beforehand with Geant4 and then introduced into the simulation at runtime. This parameterisation method successfully speeds up the most CPU consuming process, the optical photon's propagation in the LS, by a factor of 50. In the paper, the comparison of physics performance between fast and full simulation is also given.
\end{abstract}

% == keywork ==
%\end{frontmatter}

% == pacs ==

% = enable line numbers =
%\linenumbers

% two columns
\begin{multicols}{2}

%%%%%%%%%%%%%%%%%%%%%%%%%%%%%%%%%%%%%%%%%%%%%%%%%%%%%%%%%%%%%%%%%%%%%%%%%%%%%
% = introduction, the JUNO experiment =
%%%%%%%%%%%%%%%%%%%%%%%%%%%%%%%%%%%%%%%%%%%%%%%%%%%%%%%%%%%%%%%%%%%%%%%%%%%%%
\section{Introduction\label{sec:introduction}}

The Jiangmen Underground Neutrino Observatory (JUNO)\cite{Djurcic:2015vqa, An:2015jdp} is a multiple purpose neutrino experiment to determine neutrino mass hierarchy and precisely measure oscillation parameters. It is being built in Jiangmen in Southern China and is about \unit{53}{\kilo\meter} away from Yangjiang and Taishan nuclear power plants, respectively.
% figure of JUNO detector
\begin{center}
\includegraphics[width=8cm]{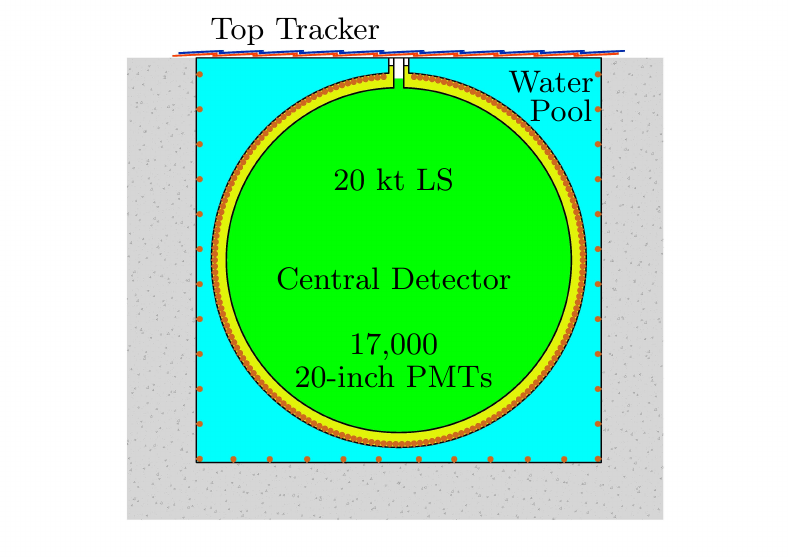}
\figcaption{\label{fig:detector_scheme} (color online) Schematic view of the JUNO detector}
\end{center}
The schematic view of the JUNO detector is shown in Fig.~\ref{fig:detector_scheme}. The most inner part is called the central detector (CD). The CD is basically an acrylic sphere filled with \unit{20}{\kilo\ton} liquid scintillator (LS) and surrounded by about 17,000 20-inch photomultiplier tubes (PMTs). An energy resolution of $3\%/\sqrt{E(\MeV)}$ is expected. Around the CD, there is a water pool to shield radioactivities. The PMTs in the water pool are used to detect the Cerenkov lights yielded by cosmic ray muons. On top of the water pool, there is a top tracker, made of plastic scintillators, to identify muon tracks.

To meet the requirement of high energy resolution for the CD, reliable and flexible Monte Carlo (MC) simulation software is a necessity, especially for optimizing detector parameters at the design stage. The simulation software has been developed  based on the Geant4 \cite{Agostinelli:2002hh} toolkit, which consists of a number of packages to manage event generators, geometry and materials, physics processes, tracking and user interfaces. Both optical parameters and physics processes have been tuned based on the experimental data obtained by the Daya Bay Neutrino Experiment\cite{An:2012eh}. The simulation software is highly integrated with the underlying software framework of SNiPER\cite{zoujh:2015sniper} so that it becomes an important component in the full data processing chain.

%%%%%%%%%%%%%%%%%%%%%%%%%%%%%%%%%%%%%%%%%%%%%%%%%%%%%%%%%%%%%%%%%%%%%%%%%%%%%
% = data processing =
%%%%%%%%%%%%%%%%%%%%%%%%%%%%%%%%%%%%%%%%%%%%%%%%%%%%%%%%%%%%%%%%%%%%%%%%%%%%%
\section{Data processing\label{sec:workflow}}
The raw data recorded by the JUNO detector need to be processed offline in order to be converted to reconstructed data, which are suitable for physics analysis. In the current phase, the raw data will be produced by the MC simulation and then are processed in the same way as the real data in the future. Following the scheme of SNiPER, various data processing steps shown in Fig.~\ref{fig:workflow} are implemented as corresponding type of algorithms, software components undertaking certain amount of calculation task. Event data pass between different algorithms via a data buffer. The event data cached  in the data buffer can be written to ROOT \cite{Brun:1997pa} files for persistent storage. The detector parameters stored in GDML \cite{Chytracek:2006be} files can be accessed  by algorithms through the detector description service in a uniform way.

% sniper framework
% figure to show the work flow
\begin{center}
\includegraphics[width=8cm]{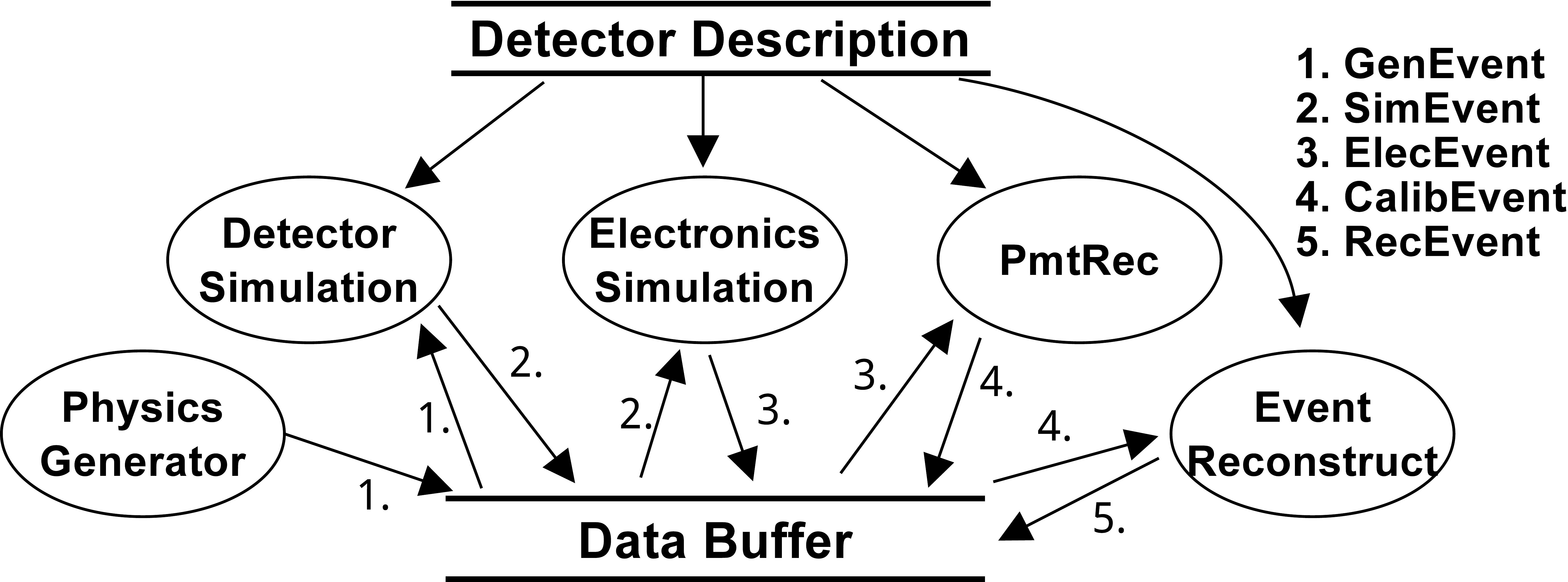}
\figcaption{\label{fig:workflow} Work flow for data processing}
\end{center}

As shown in Fig.~\ref{fig:workflow}, the physics generator generates simulated physics events such as  inverse $\beta$ decay (IBD) events, cosmic ray muon events, radioactivity events and so on. The generated GenEvent objects, which are in the format of HepMC \cite{Dobbs:2001ck}, mainly include kinematics information of the generated final-state particles. Then detector simulation loads GenEvent objects from the data buffer and starts tracking with Geant4. The information of simulated hits such as charge and time information are saved into SimEvent objects. After inputting SimEvent objects, electronics simulation performs digitization and generates ElecEvent objects containing information of waveforms with predefined sampling rate. Then the PmtRec algorithm processes the waveforms and converts them into CalibEvent objects using calibration constants. Finally, the event reconstruction algorithm reads in CalibEvent objects, performs event reconstructing and stores the produced RecEvent objects in the data buffer.

%%%%%%%%%%%%%%%%%%%%%%%%%%%%%%%%%%%%%%%%%%%%%%%%%%%%%%%%%%%%%%%%%%%%%%%%%%%%%
% = detector simulation =
%%%%%%%%%%%%%%%%%%%%%%%%%%%%%%%%%%%%%%%%%%%%%%%%%%%%%%%%%%%%%%%%%%%%%%%%%%%%%
\section{Challenge of muon simulation\label{sec:designdetsim}}
The muon induced background is one of the main backgrounds in the JUNO experiment. The mean energy of muons that penetrate the overburden is about \unit{215}{\GeV} and they reach the JUNO detector at the  rate of about \unit{3}{\Hz} \cite{An:2015jdp}. 

When a muon travels through the CD, energy is deposited in the LS and both scintillation and Cerenkov photons are emitted: the formers are emitted isotropically, while the latters are produced with a fixed angle with respect to the muon track. The ratio Cerenkov over scintillation photons is only 5\%. The yield for scintillation photons is given by the Birks' Law \cite{Birks:1951boa}. The number of scintillation photons, $dS$, as a function of the visible energy $dE_{vis}$ is given by 
\begin{equation}
    dS = A\cdot{}dE_{vis} = A\frac{dE}{1+C_{1}\delta+C_{2}\delta^{2}}~\mbox{,}~\delta=\frac{dE}{\rho dx} [\MeV\gram^{-1}\cm^2]~\mbox{,}
    \label{eqn:birkslaw}
\end{equation}
where $dE$ is the energy loss, $A$ is the light yield and $C_1$ and $C_2$ are Birks constants.

  After an optical photon is emitted in the LS, it will travel in the LS and there is a certain  possibility  for it to reach a PMT after passing through the acrylic sphere and buffering water. During the photon's propagation, it might be absorbed or scattered. At the liquid scintillator and acrylic sphere boundary, it may be refracted, or reflected, or even a total internal reflection may occur. If it reaches the photo-cathode of a PMT, a photoelectron may be generated.  The timing information of the photoelectron can be described by Eq.~\ref{eqn:hittime}, where $T_{start}$ is the initial time, $\Delta{}T_{gen}$ is the generation time of the scintillation photon, $\Delta{}T_{hit}$ is called hit time, which includes the optical photon's propagation time and the  photoelectron's generation time.
\begin{equation}
  T = T_{start} + \Delta{}T_{gen} + \Delta{}T_{hit}.
    \label{eqn:hittime}
\end{equation}

 The MC simulation of muon particles is important for physics studies, but the full simulation with Geant4 is extremely CPU consuming. To simulate a muon  with typical energy passing through the central detector, at least 50 minutes are required using a modern CPU core.   So the generation of cosmic ray muons with full simulation becomes a big challenge. The most time consuming part is the propagation of optical photons due to the huge number of photons (at the level of $10^7$).

%%%%%%%%%%%%%%%%%%%%%%%%%%%%%%%%%%%%%%%%%%%%%%%%%%%%%%%%%%%%%%%%%%%%%%%%%%%%%
%%%%%%%%%%%%%%%%%%%%%%%%%%%%%%%%%%%%%%%%%%%%%%%%%%%%%%%%%%%%%%%%%%%%%%%%%%%%%
%%%%%%%%%%%%%%%%%%%%%%%%%%%%%%%%%%%%%%%%%%%%%%%%%%%%%%%%%%%%%%%%%%%%%%%%%%%%%
%%%%%%%%%%%%%%%%%%%%%%%%%%%%%%%%%%%%%%%%%%%%%%%%%%%%%%%%%%%%%%%%%%%%%%%%%%%%%

%%%%%%%%%%%%%%%%%%%%%%%%%%%%%%%%%%%%%%%%%%%%%%%%%%%%%%%%%%%%%%%%%%%%%%%%%%%%%
% = muon simulation =
%%%%%%%%%%%%%%%%%%%%%%%%%%%%%%%%%%%%%%%%%%%%%%%%%%%%%%%%%%%%%%%%%%%%%%%%%%%%%
\section{Fast muon simulation\label{sec:fastsim}}
The scintillation photons make up the majority of optical photons generated by the muons passing through the CD. To speed up propagation process of  scintillation photons, a fast simulation employing a so-called voxel method has been implemented. Instead of the full simulation, it models the response of the PMTs  by sampling the response distributions prepared in advance with the Geant4 simulation. 

%%%%%%%%%%%%%%%%%%%%%%%%%%%%%%%%%%%%%%%%%%%%%%%%%%%%%%%%%%%%%%%%%%%%%%%%%%%%%
% == Voxel method ==
%%%%%%%%%%%%%%%%%%%%%%%%%%%%%%%%%%%%%%%%%%%%%%%%%%%%%%%%%%%%%%%%%%%%%%%%%%%%%
\subsection{Voxel method\label{subsec:voxelmethod}}
In the voxel method, the volume of the whole liquid scintillator sphere is treated as a regular grid  and a voxel represents   the volume element in the grid. By dividing the sphere into voxels, a muon  particle encounters a serial of voxels along its path when it enters the CD. In Fig.~\ref{fig:voxelmethod}, the scintillation photons generated in the shaded voxel (fired voxel) reach the PMT and generate detector response. The location relationship between the fired voxel and the PMT can uniquely be determined by a pair of variables ($R$, $\theta$), where $R$ is the radius of the voxel and $\theta$ is the spatial angle between the voxel  and the PMT. If there is a certain amount visible energy of $E_{vis}$ in a voxel, the response of the PMTs with the same ($R$, $\theta$) pair value follows the same distributions. It is obvious that the response distributions can be easily obtained by running the full simulation. So the core of the voxel method is to build the connection between the $E_{vis}$ in a voxel and the response of the PMTs.

\begin{center}
    \includegraphics[width=8.cm]{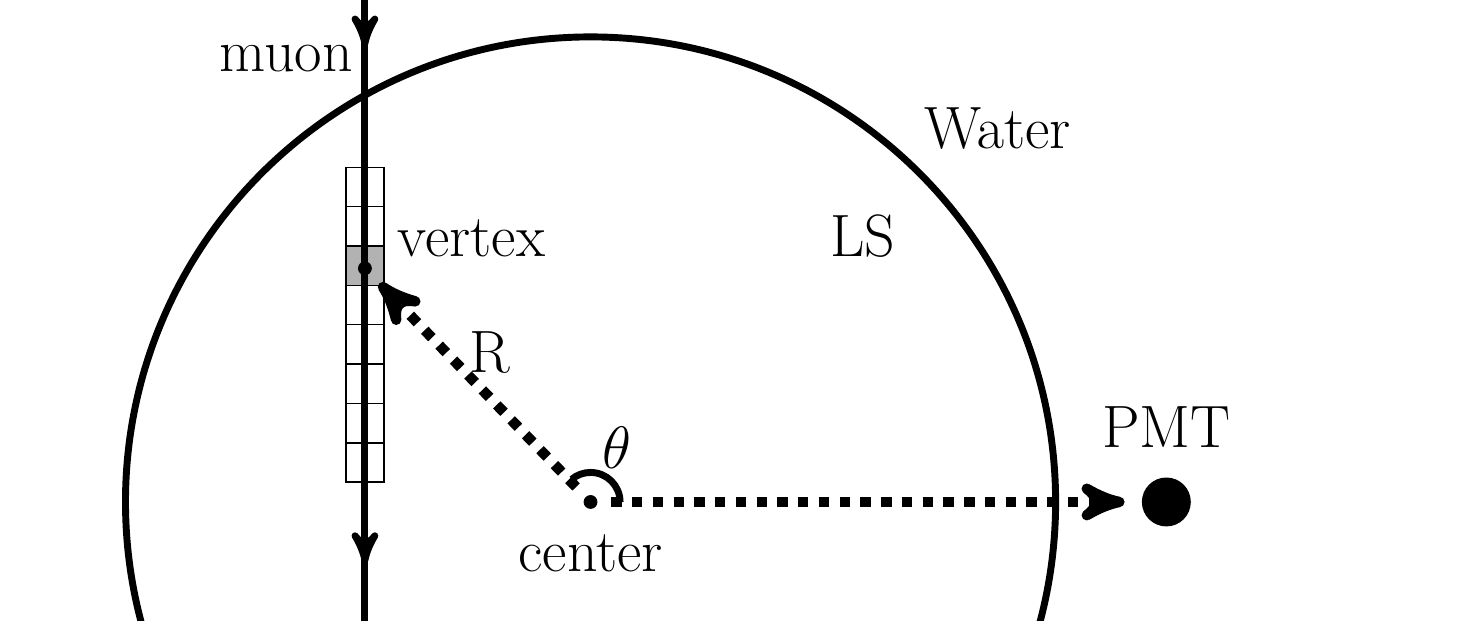}
    \figcaption{{Illustration of mapping the visible energy within a voxel to the response of the PMTs}
        \label{fig:voxelmethod}
    }
\end{center}

At the stage of preparing response distributions,  photons are produced uniformly in all LS voxels with the full simulation. Since photons with different wavelength behave differently in the detector, the emission spectrum of the LS is used  to calculate the wavelength of the photons. For each voxel, the total number of generated photons is equivalent to the light yield of \unit{1}{\MeV} visible energy, which makes it possible for response distributions to contain sufficient information for the fast simulation to simulate a physics event. Photons received by each PMT and the associated timing information are the most important data to reflect optical photon's transportation process in the LS.  So the number of photoelectrons (nPE) and hit time can be regarded as the major parameters of PMT response. 

%So for a voxel-PMT pair with ($R$, $\theta$), two histograms are generated, which are filled with nPE and hit time respectively. 
%They are further converted to photoelectrons and hit time after including more effects of the PMT.

For  simplicity, an  $R$-$\theta$ parameter space is used to represent the set of  possible combinations of ($R$, $\theta$) values, in which $R$ ranges from 0 to the inside  radius of the acrylic sphere and $\theta$ is between $0$ to $\pi$.  Both $R$ and   $\theta$ are divided into a serial of intervals, respectively. Each 2-dimensional bin in parameter space is associated with a number of voxels and PMTs. For a specific bin, two histograms, one is for nPE distribution and the other for hit time distribution, are generated by the full simulation. After traversing all the bins in the parameter space, a complete set of response distributions are obtained. Fig.~\ref{fig:hittime_r3_costheta} shows a profile of mean hit time for scintillation photons within a bin in the $R$-$\theta$ parameter space, in which the x-axis represents $R^3$, the y-axis represents $\theta$ and the z-axis indicates the mean hit time. This figure shows that vertex at the edge of detector is influenced by the total internal reflection, so that the mean hit time is larger than the normal hit time.

\begin{center}
    \includegraphics[width=8cm]{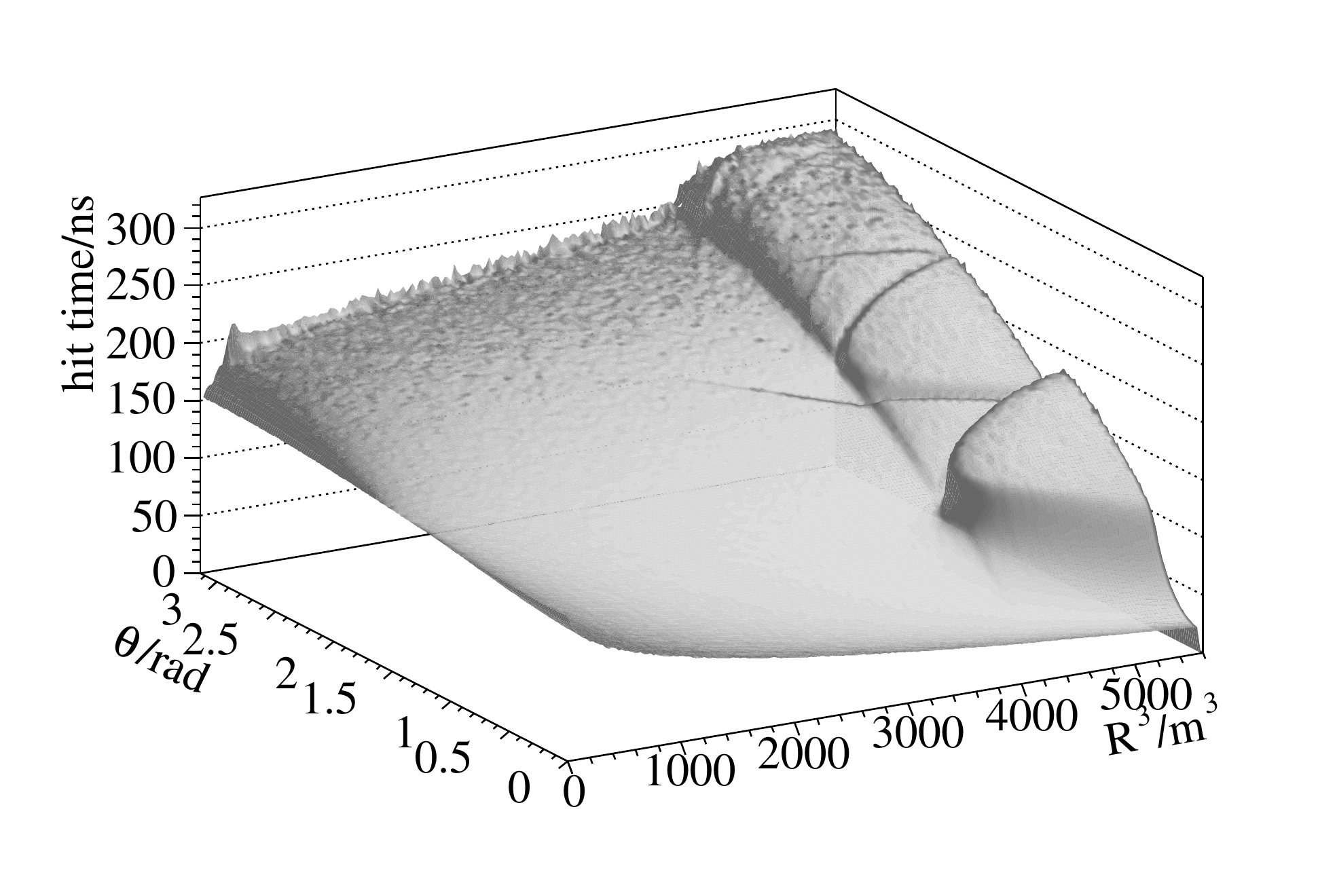}
    \figcaption{%Mean hit time vs $R^3$ and $\cos(\theta)$.
        Profile of mean hit time for scintillation photons within a bin in the $R$-$\theta$ parameter space.
        \label{fig:hittime_r3_costheta}
    }
\end{center}

% === statistics ===
%The $E_{vis}$ can be different in the same ($R$, $\theta$) pair, so the distribution of nPE is different. It's impossible to generate histograms for all range of $E_{vis}$. Assuming a photon is generated in a voxel isotropically, the probability reaching a PMT is $p$. Now, if $N_{gen}$ photons are generated, there are $N$ PEs on the PMT, where random variable $N$ follows the binomial distribution, which is shown in Eq.~\ref{eqn:bionominal}. If $N^i \sim \mathrm{B}(N_{gen}^i, p)$ and $N^j \sim \mathrm{B}(N_{gen}^j, p)$, the sum of these random variables also follows the same distribution as Eq.~\ref{eqn:sumbionominal} shows. This ensures the accuracy of sum of nPE. In our simulation, $N_{gen}$ is proportional to $E_{vis}$. So we can predicate nPE with $E_{vis}$.  
%\begin{equation}
%    N \sim \mathrm{B}(N_{gen}, p)
%    \label{eqn:bionominal}
%\end{equation}
%\begin{equation}
%    N^i+N^j \sim \mathrm{B}(N_{gen}^{i} + N_{gen}^{j}, p)
%    \label{eqn:sumbionominal}
%\end{equation}

At the stage of using the prepared response distributions to do the fast muon simulation, the procedure is as follows: for every step of muon tracking in LS, the visible energy, $E_{vis}$, is calculated according to the Birks' Law, which is the same as that in scintillation process. If the $E_{vis}$ value is none-zero, the voxel that  the tracking step  encounters is identified by the step's position and the two associated nPE and hit time histograms  are loaded. Then $E_{vis}$ is broken into an integer part (floor of $E_{vis}$, $\lfloor{}E_{vis}\rfloor$) and a fractional part. Both are used to calculate nPE, as Eq.~\ref{eqn:npeall} shows, where $N_{int}$ is the integer part and $N_{frac}$ is the fractional part. For the integer part,the nPE  histogram is sampled $\lfloor{}E_{vis}\rfloor$ times, as shown in Eq.~\ref{eqn:npeint}. For the fractional part, the nPE histogram is sampled only once to get $N_{int}^{\lfloor{}E_{vis}\rfloor}$. $N_{frac}$ follows the binomial distribution $N_{frac} \sim \mathrm{B}(N_{int}^{\lfloor{}E_{vis}\rfloor}, p)$, where the fractional part is used to calculate the probability for the PMT to be hit or not. 
\begin{equation}
    N = N_{int} + N_{frac}
    \label{eqn:npeall}
\end{equation}
\begin{equation}
    N_{int} = \sum_{j=0}^{j<\lfloor{}E_{vis}\rfloor} N_{int}^{j}
    \label{eqn:npeint}
\end{equation}
If the PMT's nPE is greater than zero, its hit time is obtained by sampling the associated hit time histogram. 

\subsection{Performance measurements\label{subsec:cmpvoxel}}
The data sample used in the physics performance studies are single $\gamma$s with the energy of \unit{1}{\MeV},  which are generated along vertical z axis with a step of  \unit{0.1}{\meter} in the LS.  To guarantee sufficient statistics,  10,000 events are generated at each position.

\newcommand{\ltcmpref}[2]{fig:fastfull_#2_#1m}
%As to the CD simulation, photons received by each PMT and the associated timing information are the most important data to reflect optical photon's transportation process in the LS. They are further converted to photoelectrons and hit time after including more effects of the PMT. So the number of photoelectrons (nPE) and hit time can be regarded as the main evaluation factors to examine the physics performance of the fast simulation.

%%%%%%%%%%%%%%%%%%%%%%%%%%%%%%%%%%%%%%%%%%%%%%%%%%%%%%%%%%%%%%%%%%%%%%%%%%%%%
% nPE and hit time distribution in one figure
%%%%%%%%%%%%%%%%%%%%%%%%%%%%%%%%%%%%%%%%%%%%%%%%%%%%%%%%%%%%%%%%%%%%%%%%%%%%%
\begin{center}
    \includegraphics[width=8cm]{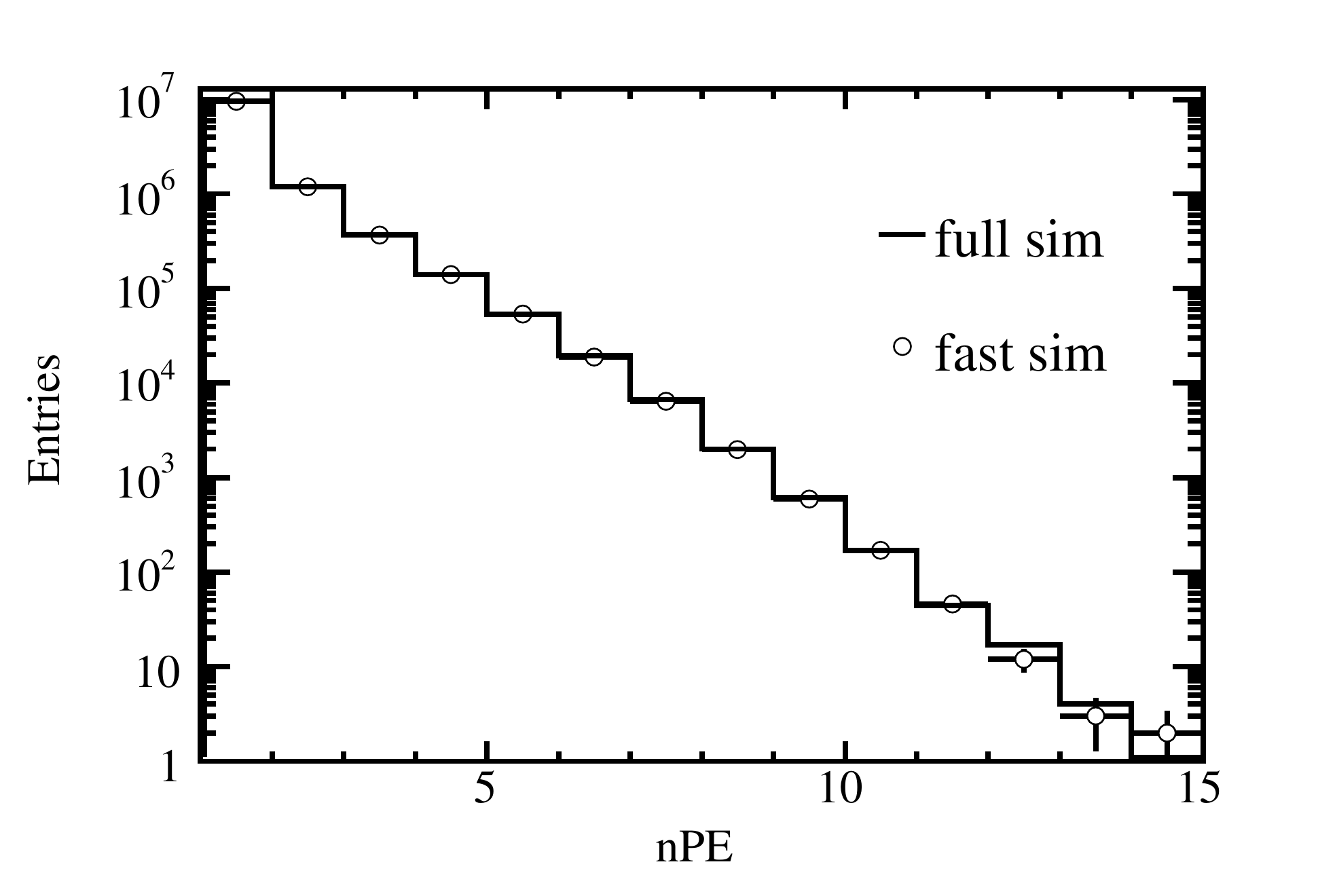}\\
    (a) nPE of PMT  \\
    \includegraphics[width=8cm]{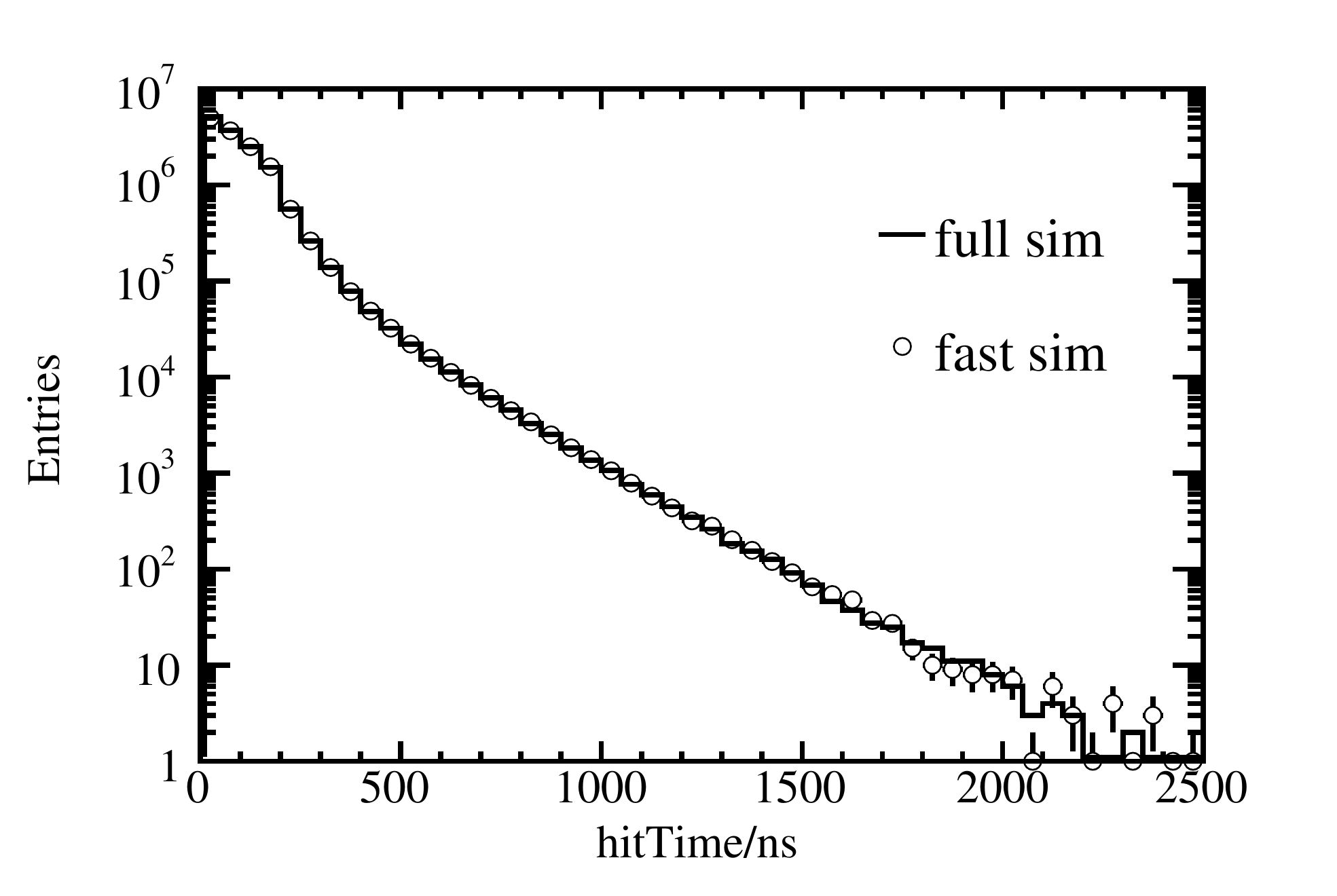}\\ % hit time only
    (b) Hit time of PMT
    \figcaption{\label{fig:cmp-npe-hittime} Comparisons of nPE and hit time between fast and full simulation using  $\gamma$s generated at z=\unit{16}{\meter}.}
\end{center}

As shown in  Fig.~\ref{fig:cmp-npe-hittime}, the distributions of nPE and hit time are compared respectively between the fast and full simulation using  $\gamma$s generated at z=\unit{16}{\meter}. It is obvious that the agreement is good for both nPE and hit time at the PMT level.  The optical photons received by a PMT should have two components: photons directly from the interaction position and reflected  photons. The closer to the acrylic sphere the $\gamma$s are generated, the more optical photons are expected to be bounced back at the barrier due to the reflection effect. For $\gamma$s generated at z=\unit{17.3}{\meter}, Fig.~\ref{fig:hittime-fixed-pos-fixed-angle} shows the hit time distribution of some selected PMTs whose $\theta$ varies from \unit{175}{\degree} to \unit{176}{\degree}. According to MC truth information of the full simulation, the first peak is caused by the direct photons, however, the second one is mainly contributed  by the reflected photons. So the fast simulation can reflect effectively the transportation of optical photons  in the LS.

\begin{center}
    \includegraphics[width=8cm]{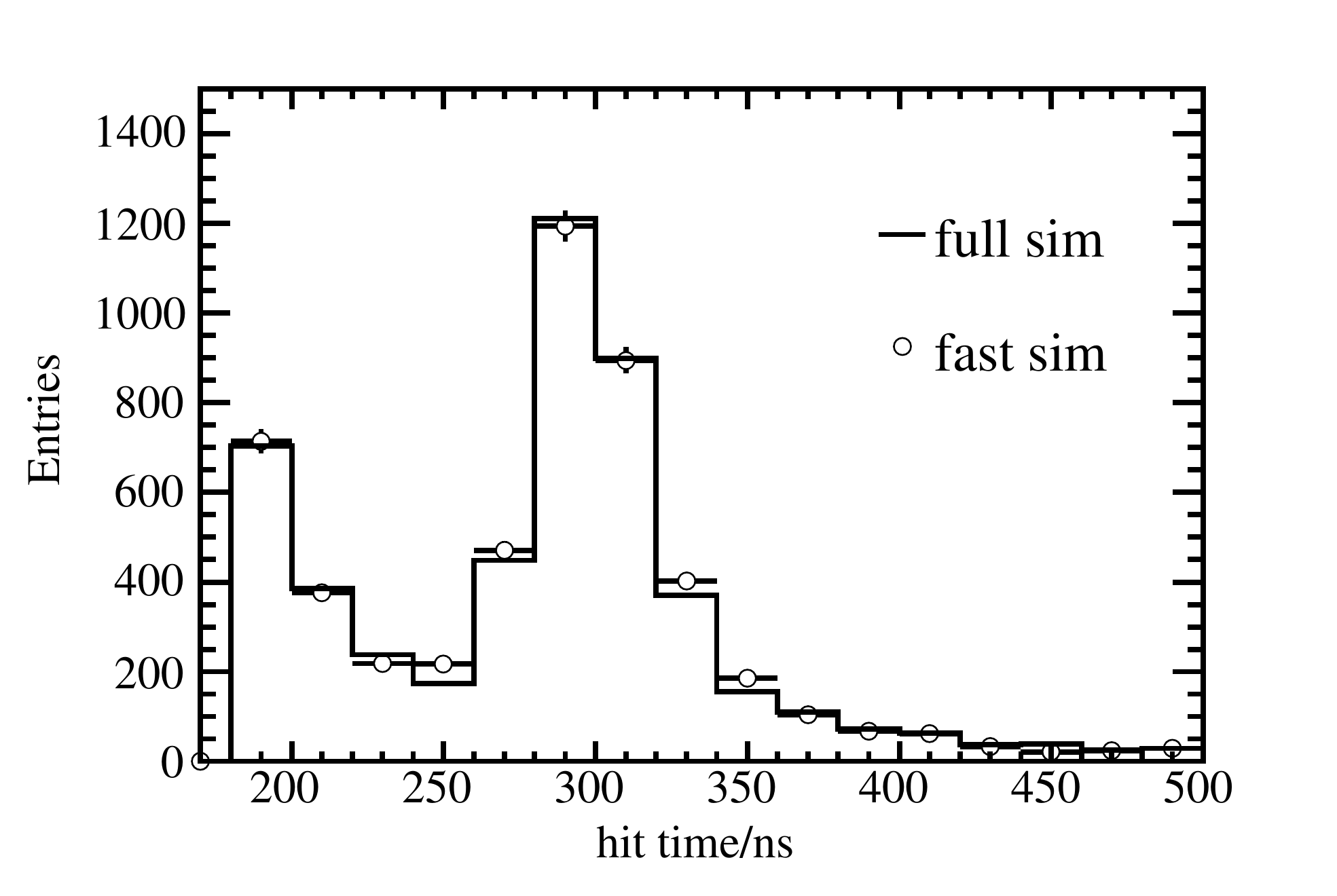}
    \figcaption{\label{fig:hittime-fixed-pos-fixed-angle}   Hit time  of the PMTs in the region  of \unit{175}{\degree}   to \unit{176}{\degree} with $\gamma$s generated at z=\unit{17.3}{\meter}.}
\end{center}

Fig.~\ref{\ltcmpref{all}{unitotalpe}} shows the average total number of photoelectrons per event (totalPE) as a function of $\gamma$'s generation position z. The difference between fast and full simulation is less than 0.4\%, which again demonstrates a good consistency between them. The totalPE increases with the $\gamma$'s generation position getting closer to to the acrylic sphere. After reaching the turning point, the totalPE drops down mainly because of the total internal reflection and energy leak. So the fast simulation can model the change of  totalPE with the vertex in a precise way.

%%%%%%%%%%%%%%%%%%%%%%%%%%%%%%%%%%%%%%%%%%%%%%%%%%%%%%%%%%%%%%%%%%%%%%%%%%%%%
% total pe
%%%%%%%%%%%%%%%%%%%%%%%%%%%%%%%%%%%%%%%%%%%%%%%%%%%%%%%%%%%%%%%%%%%%%%%%%%%%%
%\begin{center}
%    \includegraphics[width=8cm]{bundle_totalpe.eps} % fixed position
%    \figcaption{\label{fig:totalpe-alongz}Comparison of totalPE along z and with z = \unit{16}{\meter}}
%\end{center}

%\begin{center}
%    \includegraphics[width=8cm]{cmp-totalpe-16m.eps} % fixed position
%    \figcaption{\label{\ltcmpref{16}{totalpe}}Comparison of totalPE at z = \unit{16}{\meter}}
%\end{center}

\begin{center}
    \includegraphics[width=8cm]{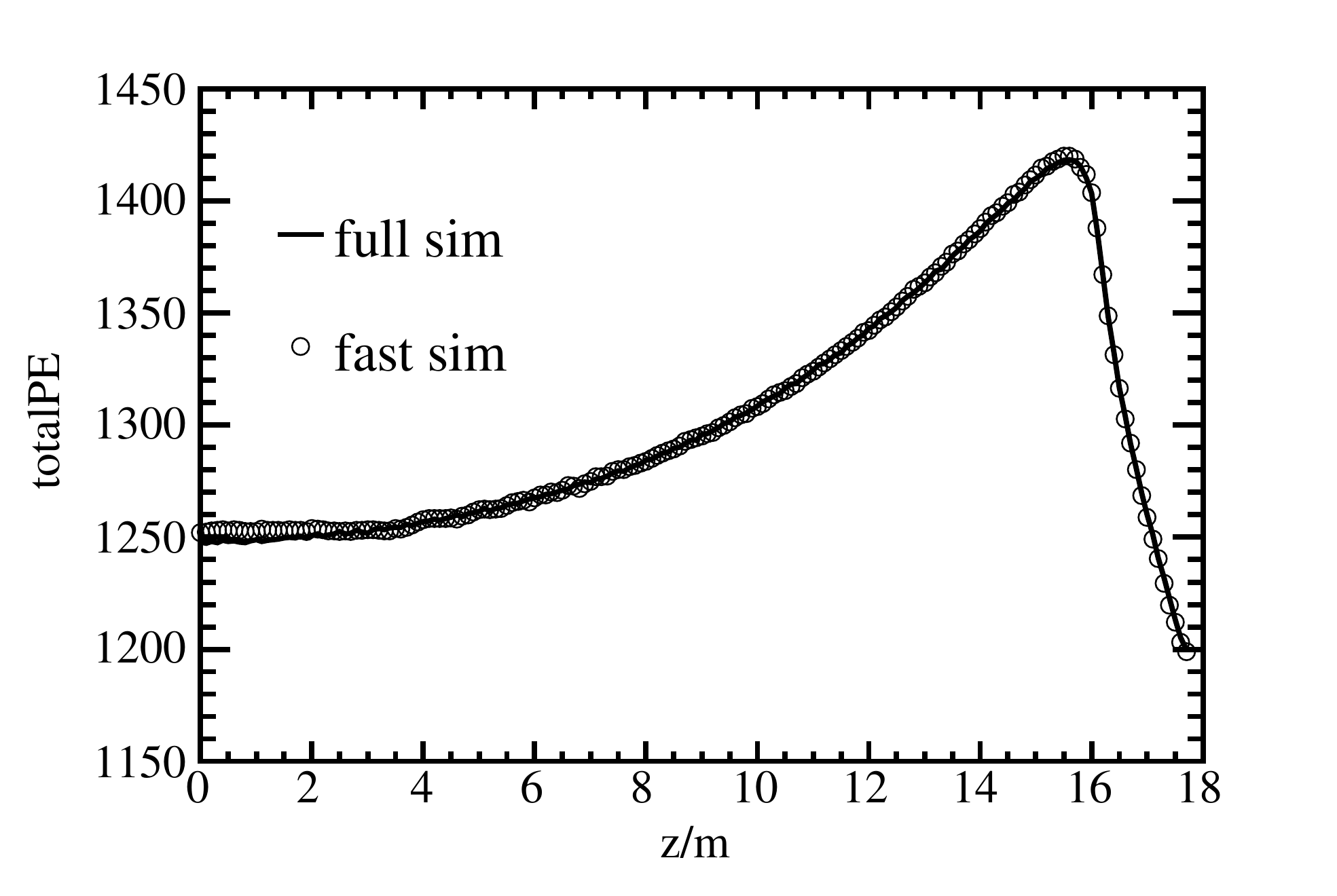} % fixed position
    \figcaption{\label{\ltcmpref{all}{unitotalpe}}Average totalPE as a function of $\gamma$'s generation position z.}
\end{center}

%%%%%%%%%%%%%%%%%%%%%%%%%%%%%%%%%%%%%%%%%%%%%%%%%%%%%%%%%%%%%%%%%%%%%%%%%%%%%
% hit time @16m
%%%%%%%%%%%%%%%%%%%%%%%%%%%%%%%%%%%%%%%%%%%%%%%%%%%%%%%%%%%%%%%%%%%%%%%%%%%%%
%\begin{center}
%    \includegraphics[width=8cm]{cmp-hittime-16m.eps} % hit time only
%    \figcaption{\label{\ltcmpref{16}{hittime}}Comparison of hit time of \unit{1}{\MeV} $\gamma$ events at z = \unit{16}{\meter}}
%\end{center}
%%%%%%%%%%%%%%%%%%%%%%%%%%%%%%%%%%%%%%%%%%%%%%%%%%%%%%%%%%%%%%%%%%%%%%%%%%%%%
% two peaks of hit time
%%%%%%%%%%%%%%%%%%%%%%%%%%%%%%%%%%%%%%%%%%%%%%%%%%%%%%%%%%%%%%%%%%%%%%%%%%%%%

When a muon particle passes through the CD, the number of scintillation photons it generates is proportional to its visible energy in the liquid scintillator. So the more energy the muon particle deposits, the more CPU time is required to simulate the generation and transportation process of scintillation photons. To make a thorough timing measurement, a broad band of energy points covering \unit{10}{\GeV}, \unit{100}{\GeV}, \unit{215}{\GeV}, \unit{500}{\GeV} and \unit{1}{\TeV} is chosen for muon particle's generation.  Ahead of timing measurements, a full simulation job is run and during job execution information at each MC simulation step are collected and stored in a data file. The prepared data file will be used by the subsequent fast simulation so that it can be compared with the full simulation at event level. The measurement of system performance is done on a blade server with the CPU Intel\textregistered{} Xeon\textregistered{} E5-2680 v3 @ 2.50GHz. In order to eliminate interferences, the CPU is exclusively used by the timing measurement job and other irrelevant applications such as system monitoring are all suspended.

%%%%%%%%%%%%%%%%%%%%%%%%%%%%%%%%%%%%%%%%%%%%%%%%%%%%%%%%%%%%%%%%%%%%%%%%%%%%%
% compare between fast and full, event by event
%%%%%%%%%%%%%%%%%%%%%%%%%%%%%%%%%%%%%%%%%%%%%%%%%%%%%%%%%%%%%%%%%%%%%%%%%%%%%
\begin{center}
    \includegraphics[width=8cm]{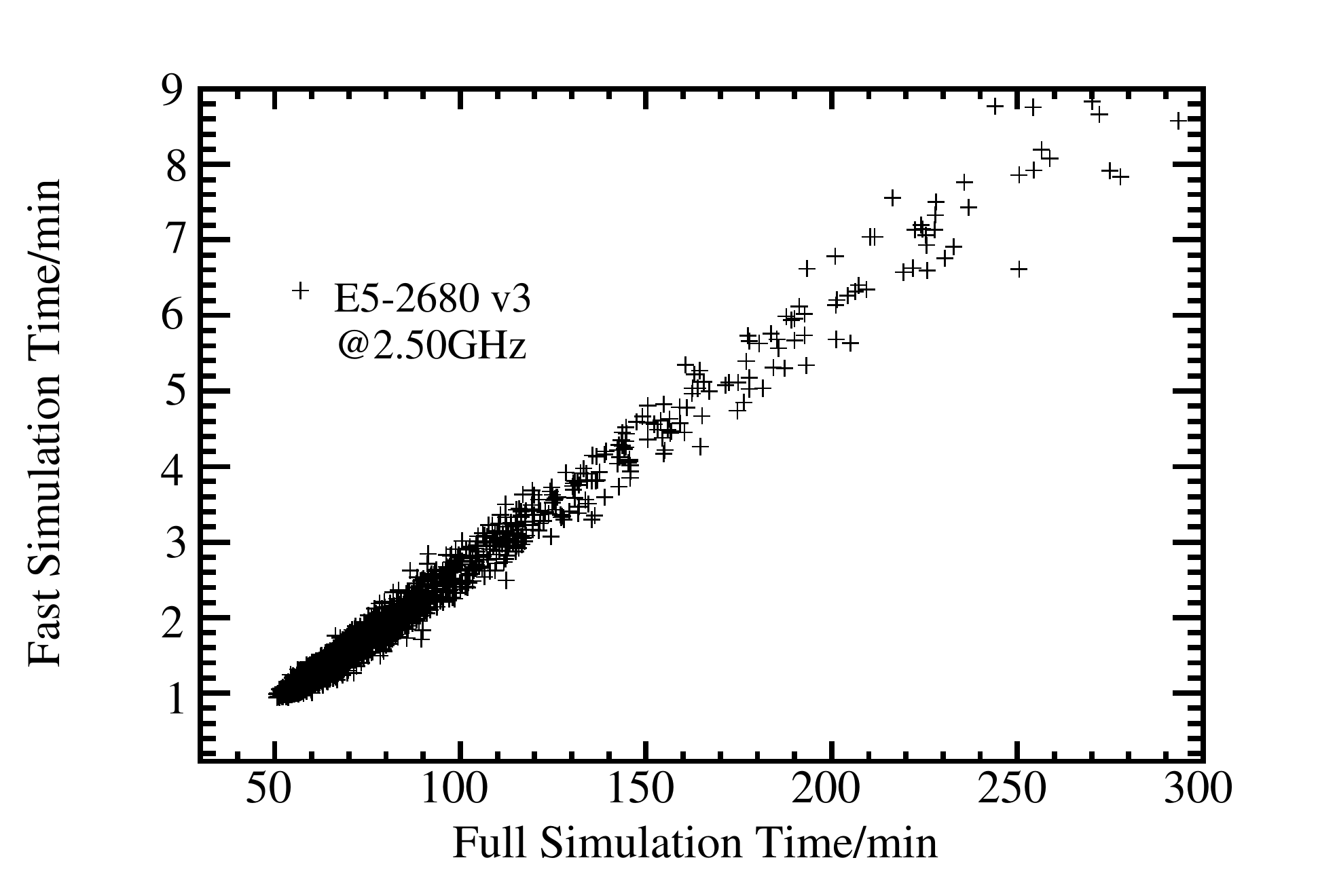}
    \figcaption{\label{fig:event-by-event-compare} The fast simulation time versus the full simulation time for muon particles with their momenta varying between  \unit{10}{\GeV} and \unit{1}{\TeV}. }
\end{center}

Fig.~\ref{fig:event-by-event-compare} is a scatter plot of execution time elapsed on generation and transportation of scintillation photons generated by muons passing through the central detector. Each cross represents the fast simulation time versus the full simulation time for a muon particle. The result shows that the elapsed simulation time scales linearly with the muon particle's visible energy and the fast simulation achieves a speedup ratio of  about 50 over the full simulation.

%\begin{center}
%    \includegraphics[width=8cm]{event_by_event_fast_vs_full_v8.eps}
%    \figcaption{\label{fig:event-by-event-speedup} Full simulation time vs speedup.}
%\end{center}

%%%%%%%%%%%%%%%%%%%%%%%%%%%%%%%%%%%%%%%%%%%%%%%%%%%%%%%%%%%%%%%%%%%%%%%%%%%%%
% == discussion ==
%%%%%%%%%%%%%%%%%%%%%%%%%%%%%%%%%%%%%%%%%%%%%%%%%%%%%%%%%%%%%%%%%%%%%%%%%%%%%
\subsection{Discussions\label{subsec:discussion}}
The JUNO CD simulation with Geant4 is largely dominated by optical photons' transportation in the liquid scintillator. Although the voxel method has been  developed to speed up the muon simulation in the central detector, it can also be extended to simulate other types of events such as IBD because the parameterization approach is related to neither event types nor event energy. 

The symmetry of detector geometry greatly reduces the number of sampling histograms used by the voxel method. But there are many factors that might break this kind of symmetry. For example supporting sticks around the acrylic sphere will block detection of optical  photons. The voxel method can handle this problem by increasing the number of sampling histograms.

It is easy to use parallel computing technique such as CUDA \cite{CUDA} to further accelerate execution of the voxel method. Unlike the full simulation application running on GPU such as Chroma \cite{chroma}, the parallel implementation of voxel method can only execute histogram sampling with GPUs, which avoids the complexity of geometry translation. 

%%%%%%%%%%%%%%%%%%%%%%%%%%%%%%%%%%%%%%%%%%%%%%%%%%%%%%%%%%%%%%%%%%%%%%%%%%%%%
% = data production and simulation performance =
%%%%%%%%%%%%%%%%%%%%%%%%%%%%%%%%%%%%%%%%%%%%%%%%%%%%%%%%%%%%%%%%%%%%%%%%%%%%%
%\section{Physics performance}

%%%%%%%%%%%%%%%%%%%%%%%%%%%%%%%%%%%%%%%%%%%%%%%%%%%%%%%%%%%%%%%%%%%%%%%%%%%%%
% = summary and conclusion =
%%%%%%%%%%%%%%%%%%%%%%%%%%%%%%%%%%%%%%%%%%%%%%%%%%%%%%%%%%%%%%%%%%%%%%%%%%%%%
\section{Conclusions}
The fast muon simulation using the voxel method has been implemented in the JUNO's offline software framework. The applied method is to generate the response of the PMTs beforehand, for photons produced in different voxels in the CD with Geant4 and then to introduce the response into the fast simulation by sampling the response histograms at runtime. 

The timing measurement shows that the fast simulation has obtained a speedup ratio of 50 in CPU execution time compared to the full simulation with Geant4. The physics performance has also been examined and validation results show that the agreement between the fast and full simulation is good and no significant discrepancies have been found. 

%The fast muon simulation in the central detector  meets the requirements of the JUNO experiment. 

%%%%%%%%%%%%%%%%%%%%%%%%%%%%%%%%%%%%%%%%%%%%%%%%%%%%%%%%%%%%%%%%%%%%%%%%%%%%%
% = Acknowledgments =
%%%%%%%%%%%%%%%%%%%%%%%%%%%%%%%%%%%%%%%%%%%%%%%%%%%%%%%%%%%%%%%%%%%%%%%%%%%%%
%\section*{Acknowledgments}
\ifpaperthanks
\acknowledgments{
This work is supported by the Strategic Priority Research Program of the Chinese Academy of Sciences, Grant No. XDA10010900.
}
\fi

\end{multicols}
%%%%%%%%%%%%%%%%%%%%%%%%%%%%%%%%%%%%%%%%%%%%%%%%%%%%%%%%%%%%%%%%%%%%%%%%%%%%%
% = reference =
%%%%%%%%%%%%%%%%%%%%%%%%%%%%%%%%%%%%%%%%%%%%%%%%%%%%%%%%%%%%%%%%%%%%%%%%%%%%%
% a horizon line here
\vspace{-1mm}
\centerline{\rule{80mm}{0.1pt}}
\vspace{2mm}

\begin{multicols}{2}
%\section*{References}
%\bibliography{mybibfile}

\end{multicols}
\clearpage

\ifpapercjk
\end{CJK*}
\fi
\end{document}